{{}} 

\documentclass[12pt]{iopart}
\usepackage{graphicx}
\usepackage{placeins}
\usepackage{float}
\usepackage{setspace}
\usepackage{todonotes}
\usepackage[backend=biber,style=numeric,sorting=none]{biblatex}
\usepackage{svg}
\usepackage{hyperref}
\hypersetup{pdfcreator={Me}}
\hypersetup{pdfproducer={Me}}
\begin{document}

\title[3D FE-Based Multiphysics Simulation of a SMAHC actuator module]{3D Finite Element-Based Multiphysics Simulation of a Shape Memory Alloy Hybrid Composite Module}
\author{L Handl, M Kaiser, M Duhovic 
and M Gurka}
\address{Leibniz-Institut für Verbundwerstoffe GmbH, Erwin-Schrödinger-Straße 58, Kaiserslautern 67663, Rheinland-Pfalz, Germany}

\ead{miro.duhovic@leibniz-ivw.de}
\begin{abstract}
Shape adaptive shape memory alloy hybrid composites (SMAHCs) are composites that incorporate shape memory alloys (SMAs) to realize reversible and continuous shape transformation. Despite the availability of numerous analytical and finite element models for predicting the transient response of SMAHCs, many approaches still exhibit limitations with respect to the electro-thermomechanical coupling and comprehensive experimental validation. Therefore, this paper presents a coupled, multiphysics, three-dimensional finite element (FE) approach for the simulation of a SMAHC actuator, integrating mechanical, thermal, and electromagnetic solvers in the Finite Element Code ANSYS LS-DYNA. The proposed approach employs a micromechanical constitutive model by Kelly et al., implemented in ANSYS LS-DYNA, to accurately capture the complex thermomechanical phase transformation of SMAs. A key feature of the model is the ability to prescribe a defined martensitic pre-strain through a preceding simulation step, in which an initially scaled SMA wire is mechanically loaded and stretched to its nominal length. This procedure enables partial detwinning of the martensitic microstructure and provides a physically motivated initialization of the material state. Joule heating of the SMA wires, as well as varying mechanical loads and ambient temperature conditions, are explicitly considered. The simulation results are validated against experimental data and a fully coupled transient staggered scheme model (SSM) to assess the predictive capability of the 3D approach. The results show good qualitative agreement, reproducing the characteristic hysteresis of actuator deflection as a function of temperature. Quantitatively, the predicted deflections are of the correct order of magnitude, although marginally outside the 95 \% experimental confidence interval. Overall, a consistent trend between simulation and experiment is observed, giving rise to possibility of simulating more complex 3D SMAHC systems.

\end{abstract}

\vspace{2pc}

\submitto{\SMS}
\begin{doublespace}  
\newpage

\section{Introduction}
\label{Introduction}

\begin{figure}[b]
	\centering
	\includegraphics[width=1\linewidth]{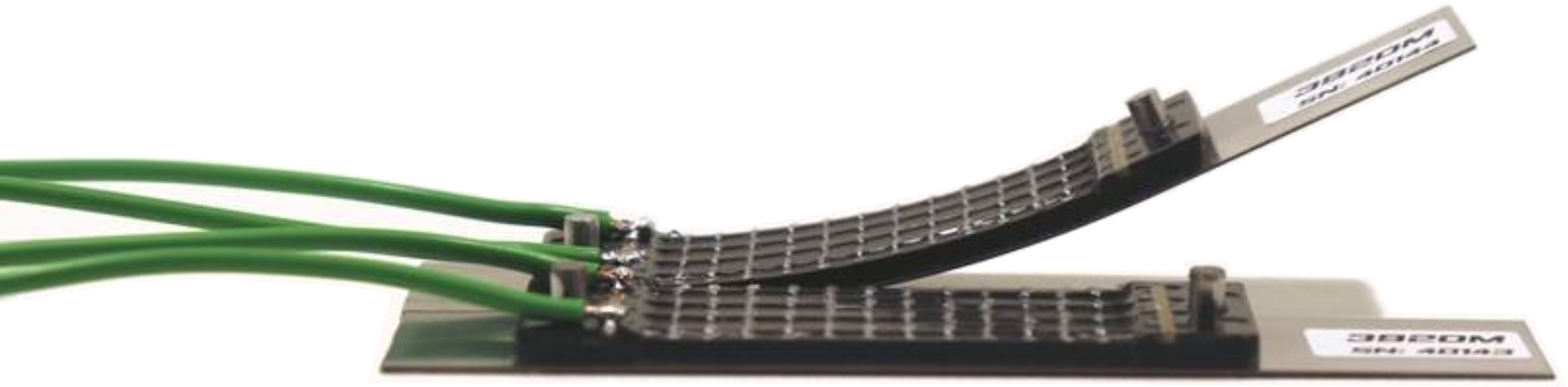}
	\caption{Illustration of an SMAHC actuator module from CompActive GmbH \cite{Compactive}.}
	\label{fig: Actuator}
\end{figure}

Actuator systems based on shape memory alloys (SMA) are becoming increasingly popular, especially as part of so called shape memory alloy hybrid composites (SMAHC), shown in Figure \ref{fig: Actuator}. Particularly as active spoilers for aerodynamic optimization, such as smart rim openings \cite{Compactive}, shape adaptive trailing edges \cite{kaiserAirfoilTrailingEdge2022}, deployable vortex generators \cite{nissleAdaptiveVortexGenerators2018}, adaptive chevrons \cite{turnerDesignFabricationTesting2006} as well grippers in robotics \cite{Compactive} and as actuators with reduced size and complexity in consumer level devices (e.g. contactless bins, automatic locks \cite{Compactive}). These actuators have the ability to significantly change their shape in response to a thermal and / or stress stimulus due to a thermoelastic phase transformation from a martensitic to an austenitic crystal structure via embedded SMA wires \cite{lagoudasShapeMemoryAlloys2008}.\\
The concept of embedding SMA wires in a polymer composite material was first introduced in the early 1990s for controlling vibrations or buckling  \cite{bazActiveVibrationControl1990, rogersActiveVibrationStructural1990}. By positioning the wire outside the neutral axis of the SMAHC, it can be used for active shape control, triggered e.g. by Joule heating, through out-of-plane bending \cite{chaudhryBendingShapeControl1991, daghiaShapeMemoryAlloy2008}. To support the development process of SMA based applications, analytical or simplified simulation models have been developed. In terms of thermomechanical modeling, the effective coefficient of thermal expansion model (ECTM) as proposed by Turner et al. \cite{turnerNewThermoelasticModel2000} is still state of the art today. This is a finite element (FE) formulation for SMA under thermal and mechanical loading for constrained, restrained and free recovery. It was implemented in the commercially available multiphysics FE modeling software ABAQUS, and its functionality for accurately predicting thermomechanical material response was demonstrated \cite{turnerAnalysisSMAHybrid2007}. This supported the development of many applications, e.g. adaptive chevrons \cite{turnerDesignFabricationTesting2006, liuThermomechanicalBehaviorsFunctional2018} or adaptive vortex generators \cite{nissleAdaptiveVortexGenerators2018}. However, the model does not take into account transient response or the electro-thermomechanical coupling. In addition, various other models have been implemented in different commercially available software systems such as COMSOL \cite{yangFiniteElementAnalysis2009}, ABAQUS \cite{turnerNewThermoelasticModel2000}, ANSYS LS-DYNA \cite{kapuriaImprovingHydrodynamicEfficiency2018}, and Matlab \cite{keshtkarMathematicalModelingFiberElastomer2020}. However, only a few consider the thermal and electrical conductivity of the embedded SMA wire or take into account essential factors such as ambient temperature or applied mechanical loads. Moreover, many of these models are often not sufficiently validated. A recent review on multiphysical modeling of SMA \cite{kaiserActiveHybridStructures2026} reveals that the approaches of Tanaka \cite{tanakaTHERMOMECHANICALSKETCHSHAPE1986}, Brinson \cite{brinsonFiniteElementAnalysis2005}, Liang and Rogers \cite{keshtkarMathematicalModelingFiberElastomer2020}, Auricchio \cite{lohseExperimentalNumericalAnalysis2022}, Müller-Achenbach-Sellecke \cite{papenfussSimulationControlSMA1999}, and Lagoudas \cite{jungNumericalSimulationVerification2013} still form the basis for the implementation of SMA-modeling considering FE-based approaches. An alternative approach, which focuses specifically on the modeling of SMAHC, uses an analytical formulation based on the theory of Tanaka, but considers the thermal coupling of the SMA wires to the composite surrounding in a more comprehensive way. As it has been validated experimentally with good agreement, this emphasizes the importance of this detail \cite{kaiserExperimentallyCharacterizationTheoretical2023}.
However, this model is a standalone solution, implemented in Python and its usage could be increased by integrating it into three-dimensional (3D) FE models. Rodino et al. \cite{rodinoMultiphysicsModelingOptimization2024} also approached the modeling of SMAHC, but in this case utilizing FE analysis using the COMSOL software and a material model based on minimizing Gibbs free energy with the formulation of Boyd and Lagoudas \cite{boydThermodynamicalConstitutiveModel1996}.\\ 
In order to support the design process of material integrated actuators, such as SMAHC within the context of computer-aided engineering (CAE) on a broader basis than today, sufficiently simple but reliable models are required which -- in contrast to Turners widely accepted approach -- also precisely describe the dynamic thermomechanical response of the SMA material under various boundary conditions.
If the exact 3D geometry of the application is important, multiphysics FE modeling today provides a useful engineering tool to describe and help to predict a wide range of physical phenomena. However, until today, there is no comprehensive and satisfactorily validated method that simulates a detailed 3D FE model of an SMAHC, taking into account boundary conditions such as ambient temperatures and applied loads, while also accounting for full electro-thermo-mechanical coupling. In the present study we therefore investigated a the capability of a new material model \cite{kellyMicromechanicsinspiredConstitutiveModel2016} that was recently implemented in the FE-based simulation software ANSYS LS-DYNA \cite{Homea} to model not only the response of SMA, namely its pseudoplasticity, superelasticity and one way shape memory effect, but also a complete SMAHC, exhibiting also external two way shape memory effect.  A detailed 3D-model of an commercially available SMAHC actuator was created and fully coupled multiphysics simulations were carried out. The results were then compared with a recently published, fully coupled staggered scheme model (SSM) \cite{kaiserExperimentallyCharacterizationTheoretical2023} of which the Python source code is available under MIT license \cite{kaima92KaiMa92Sa_smahc_py2023}, together with a comprehensive experimental dataset \cite{kaiserSMAHCCharacterizationInfluence2023}.

\section{Methods and Model}\label{Methoden}

This study utilizes the integrated coupled multiphysics (mechanical, thermal, and electromagnetic (EM)) solvers of ANSYS LS-DYNA \cite{Homea} to simulate the thermomechanical response of an SMAHC actuator within a 3D FE environment. The FE model of the SMAHC takes advantage of a recent implementation of the material model based on \cite{kellyMicromechanicsinspiredConstitutiveModel2016}, which mathematically describes the microscopic material behaviour of SMAs, taking into account the initiation, saturation and growth kinetics of the individual Martensite and Austenite phases. With the help of the aforementioned software ANSYS LS-DYNA, the experimental conditions can then be simulated in a parametrized way, using keywords to define material properties and account for varying loading and boundary conditions. The geometry of the FE model investigated here represents a commercially available SMAHC actuator (Type A3950, CompActive GmbH, Germany \cite{Datenblaetter}) with different interlayer thicknesses, attached to steel plates. In the real SMAHC actuator, the embedded SMA wires exhibit a partially detwinned martensitic microstructure in the neutral state at room temperature. However, the material card used to define the SMA parameters in the preprocessor of the employed simulation software ANSYS LS-DYNA does not allow for the specification of the initial phase state. Instead, the material is always initialized in a fully austenitic state. Therefore, an additional simulation step was introduced prior to the actual actuator simulation in order to establish a physically representative initial state corresponding to the real actuator (see Section \ref{Modellaufbau}). To validate the simulation results and enable comparison with other modeling approaches, theoretical data from a one dimensional SSM \cite{kaiserExperimentallyCharacterizationTheoretical2023} and experimental data for the aforementioned actuator module A3950 were used \cite{kaiserSMAHCCharacterizationInfluence2023}.
As some of the details, regarding design and materials used in the commercial actuators were not available from the manufacturer, data was approximated from previous work \cite{kaiserExperimentallyCharacterizationTheoretical2023} or approximated from literature where mentioned. 

\subsection{Material model}\label{Materialmdel}

The mathematical formulation of the material model is comprehensively described in \cite{kellyMicromechanicsinspiredConstitutiveModel2016} and \cite{karlssonMAT_291NewMicromechanicsinspired2019}. It assumes a polycrystalline SMA specimen, in which each (FE-) material point corresponds to a material volume with many grains, with a possible Martensite sub-microstructure within each grain. The growth kinetics and saturation of the phase transformation from Martensite to Austenite and vice versa are resolved at the microscale. Initiation and saturation are treated as fundamentally distinct processes and are therefore represented by separate potential energy contributions in Eq. \ref{eq: 1}, each formulated as a function of internal state variables, namely the Martensite volume fraction $\lambda$ ($0\leq \lambda \leq 1$) and the nominal Martensite strain tensor $\varepsilon_m$. The model is then based on minimizing the Helmholtz free energy $W$, according to \cite{kellyMicromechanicsinspiredConstitutiveModel2016} where the following equation applies:
\begin{eqnarray}\label{eq: 1}
\fl
    W(\varepsilon, \lambda, \varepsilon_{m}, T) = &\space \underbrace{\frac{1}{2}\sigma:\varepsilon}_{Elastic\ energy} + \underbrace{\lambda L \frac{T-T_{C}}{T_{C}}}_{Excess\ chemical\ energy\ Martensite} - \underbrace{c(\lambda)T\ln{(\frac{T}{T_{C}})}}_{Thermal\ energy}\nonumber\\& + \underbrace{\lambda G_{I}(\varepsilon_{m})}_{Initiation\ energy} + \underbrace{G_{s}(\lambda\varepsilon_{m})}_{Saturation\ enery} + \underbrace{G_{\lambda}(\lambda)}_{Growth\ energy}
\end{eqnarray}
The other state variables are the temperature $T$, the total strain $\varepsilon$ and the elastic strain $\varepsilon_e$.\\
The total strain
\begin{eqnarray}
\varepsilon = \varepsilon_e + \lambda\varepsilon_m
\end{eqnarray} 
is composed of the elastic and martensitic strain.\\
The first term in in Eq. \ref{eq: 1} is the elastic energy density, assuming the material to be isotropically elastic:
\begin{eqnarray}
\sigma = C(\lambda) : \varepsilon_e
\end{eqnarray}
where
\begin{eqnarray}
C(\lambda) = \lambda C_m + (1 - \lambda)C_a
\end{eqnarray}
is the effective stiffness resulting from the superposition of the martensitic stiffness $C_m$  and austenitic stiffness $C_a$.\\
The second term describes the excess chemical energy density of Martensite, whereby the model is limited to a moderate temperature range close to the phase transformation-temperature $T_C$, as the term is linearized around $T_C$. The latent heat released during the phase-transformation is represented by the variable $L$.\\ The third term describes the thermal energy density. The resulting heat capacity $c(\lambda)$ is obtained from the combination of the heat capacities of Martensite $c_m$ and Austenite $c_a$ according to their phase fraction
\begin{eqnarray}
c(\lambda) = \lambda c_m + (1 - \lambda)c_a
\end{eqnarray}\\ 
Finally, the last three terms describe the energy densities for initiation, saturation, and growth of the martensitic phase fraction. For a more comprehensive description of the mathematical modeling of the growth kinetic on the microscopic length scale please refer to \cite{kellyMicromechanicsinspiredConstitutiveModel2016} and  \cite{karlssonMAT_291NewMicromechanicsinspired2019}.

\subsection{Geometric structure of the FE-model}\label{Modellaufbau}
\begin{figure}[h]
	\centering
	\includegraphics[width=1\linewidth]{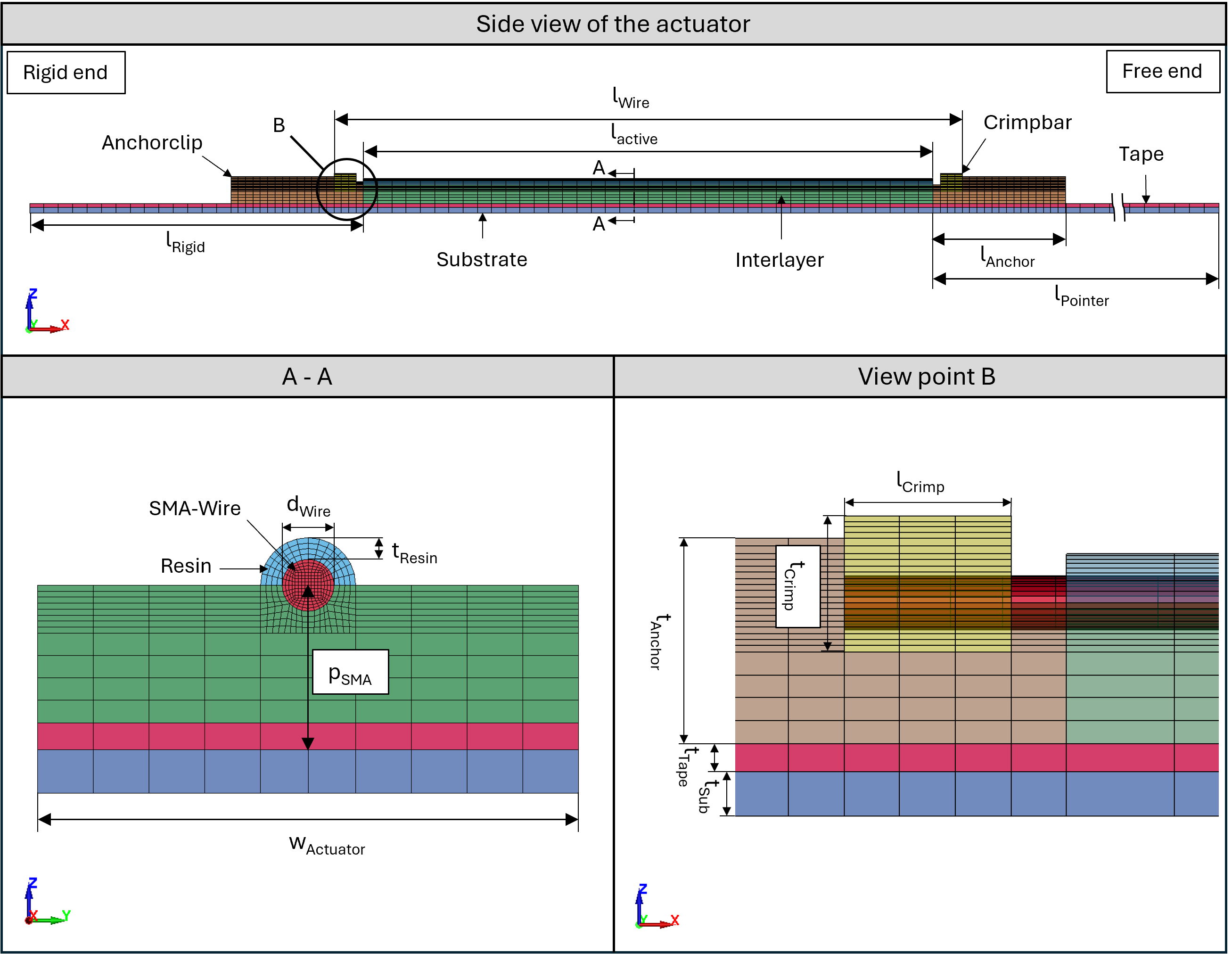}
	\caption{Geometric features of the mesh model of the simulated SMAHC actuator, Type C. Top: Side view along the active length of the SMAHC.\\
    Bottom left:Cross-section of the SMAHC at the center of the active length.\\
    Bottom right: Close-up of the SMA wire mount.}
	\label{fig: Ansichten_Aktuator}
\end{figure}

\begin{figure}[h]
	\centering
	\includegraphics[width=1\linewidth]{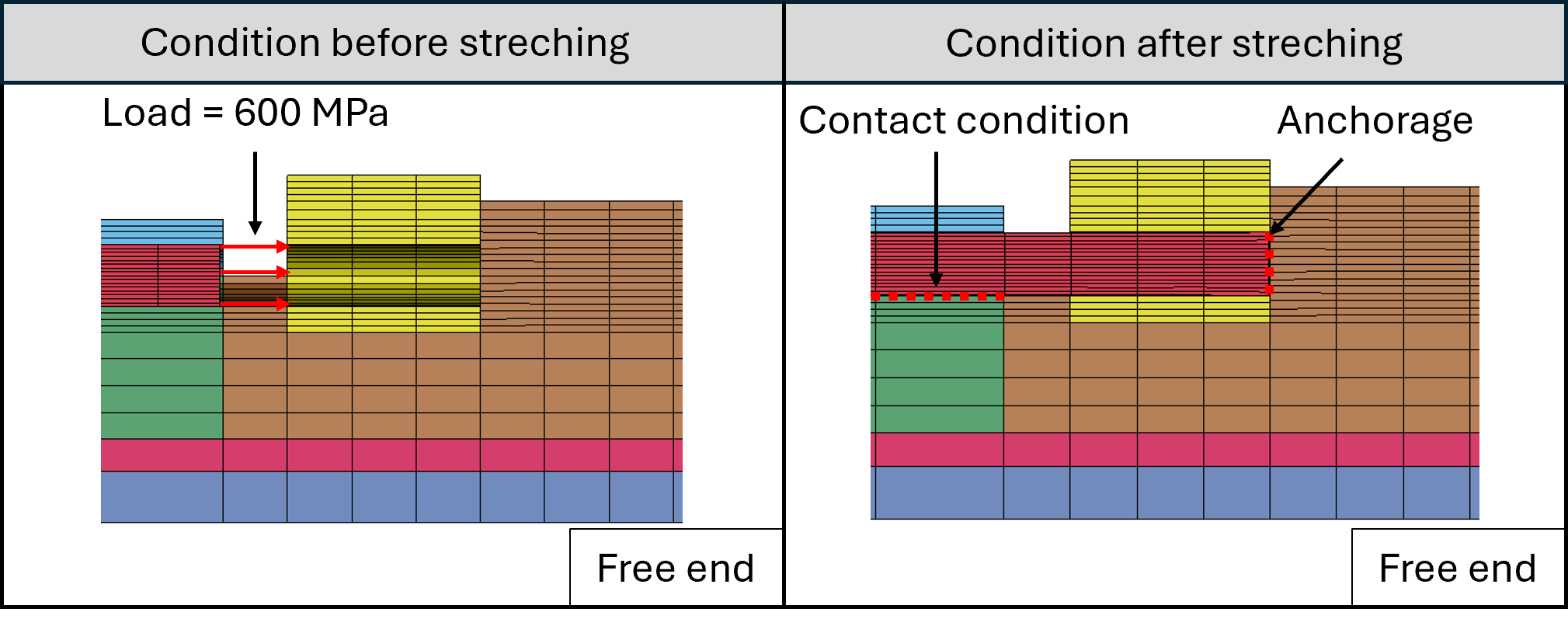}
	\caption{Schematic representation of the pre-stretching process at the free end of the actuator specifying the initial degree of detwining of Martensite to Austenite.}
	\label{fig: Vordehnung}
\end{figure}

Within the experimental test series used to validate the  model, there are four actuator types that differ in terms of geometric features, namely, the distance between the SMA wires and the elastic substrate ($p_{SMA}$) and the thickness of the elastic substrate itself. The four actuator types investigated are, depending on $p_{SMA}$, referred to below as Type A, Type B, Type C, and Type D (see also Table \ref{GeoParameter}). Apart from this difference, the actuators examined are identical. The design of such an actuator is representatively shown in Figure \ref{fig: Ansichten_Aktuator} using the FE mesh model.
\\
The associated geometric parameters are listed in \ref{GeoParameter}. Symmetries were exploited during creation of the model in order to simplify the model as much as possible. Therefore it only shows a section that contains a single SMA wire. The simulated actuator essentially consists of an SMA wire, which contracts when thermally activated and is thus responsible for the curved deformation of the actuator, an elastic substrate that serves as a return spring, a soft interlayer connecting the SMA wire and the substrate, and anchors that secure the wires to the substrate. All materials used in this model were assumed to be linearly elastic, with the exception of the SMA wire, which is modeled using the material model described in Section \ref{Materialmdel}. The parameters associated with the individual components can be found in table \ref{MechParameter} (mechanical parameters) and table \ref{ThermalParameter} (thermal parameters). The parameters of the SMA wire, which were taken from \cite{kaiserExperimentallyCharacterizationTheoretical2023, karlssonMAT_291NewMicromechanicsinspired2019}, are shown in Table \ref{MechParameterSMA}. Solid elements were used to generate the model mesh, as the material model is currently limited to this element type. Only hexahedral elements were used, as they offer better numerical accuracy than tetrahedral elements for the same element size, can represent stresses and deformations more linearly and uniformly, and can generally represent larger deformations without numerical instabilities  \cite{schneiderLargeScaleComparisonTetrahedral2022}. A total of 32,360 solid hexahedron elements were used to represent the actuator model. Within the selected mesh resolution as well as the used settings and hardware (SMP double, four cores used, Intel Core i7 11700K, 64 GB DDR4 RAM), each simulation took eight to ten hours to complete.\\ 
Since only Joule heating was modeled, a simplified form of the EM solver was used, which calculates only resistance and no induction effects. This EM solver was coupled with the structural thermal solver by adding Ohmic heating as an additional heat source. The EM fields were solved using a FE method for the conductors and a boundary element method (BEM) for the surrounding air/insulators. Therefore no air mesh is required.\\
The simulation is divided into two steps. First, the SMA wire, which was initially scaled down accordingly to compensate for longitudinal elongation and transverse contraction, is pre-stretched at an initial ambient temperature of 23 °C to its nominal length of the physical actuator model. Therefore, a load is applied to the free end of the SMA wire, which is anchored in the model via time-dependent contact conditions once the desired pre-stretch has been reached (see Figure \ref{fig: Vordehnung}). Due to the initial ambient temperature (23 °C), the wire first transforms from initial Austenite into twinned Martensite and is then partially converted into a detwinned martensitic state by the subsequent stretching. This step is necessary because the contraction potential of the wire is based on the phase transformation of the detwinned Martensite to Austenite, and the degree of detwining of the Martensite cannot be specified manually within the initial state of the simulation model. Once this step is complete, various deflection scenarios can be simulated in the second step. Joule heating of the SMA wires, as well as varying mechanical loads and the convection with the surrounding air can be simulated using corresponding boundary conditions already implemented in LS-DYNA. The most important input variables that significantly influence the deflection are the electric current, the ambient temperature, and the mechanical load applied to the free end. These parameters can be specified in the simulation's start file. 

\subsection{Experimental setup }
\label{Experimentellen Daten}

The experimental data used in this study to validate the simulation is published in \cite{kaiserExperimentallyCharacterizationTheoretical2023} and the test method is described in detail in \cite{kaiserTransientElectrothermomechanicalModeling2023}. In this study, a test environment was set up to characterize SMAHC actuation performance under different ambient temperature and load scenarios. Experiments started by fixing an SMAHC actuator on a sample holder at one end, referred to as fixed end. The opposite side of the SMAHC actuator was referred to as free end. The free end also included a so called pointer to leverage the deflection. Weights were attached to the free end to test SMAHC actuation performance under external mechanical load scenarios. The deflection of the actuator in the z-direction was measured along the x-axis at two points on the pointer using two laser triangulation sensors. To record the temperature, an infrared camera was used. A high-resolution monochrome camera allowed for digital image correlation for local deflection measurement. Electrical power was supplied with controlled current flux and the voltage drop over the SMAHCs wire was measured. Ambient temperature was controlled using a temperature chamber. The stop criteria for activation was either when the measured temperature of the resin layer surrounding the wire exceeded 403 K or when Joule heating was ongoing for more than 100 s. For more detail on the experimental procedure please refer to \cite{kaiserExperimentallyCharacterizationTheoretical2023}.

\section{Results and Discussion}
\label{Results}

\begin{figure}[h]
	\centering
	\includegraphics[width=1\linewidth]{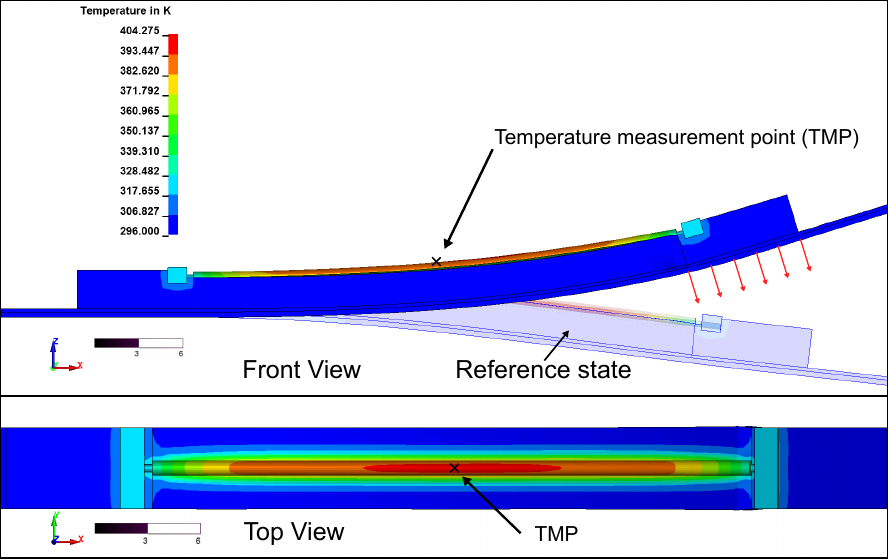}
	\caption{Deformation of the FE model at different simulation time steps (maximum and minimum actuator deflection under a load of 39.6 N), including the temperature distribution shown as isocontours.}
	\label{fig: Auslenkung}
\end{figure}

\begin{figure}[h]
	\centering
	\includegraphics[width=1\linewidth]{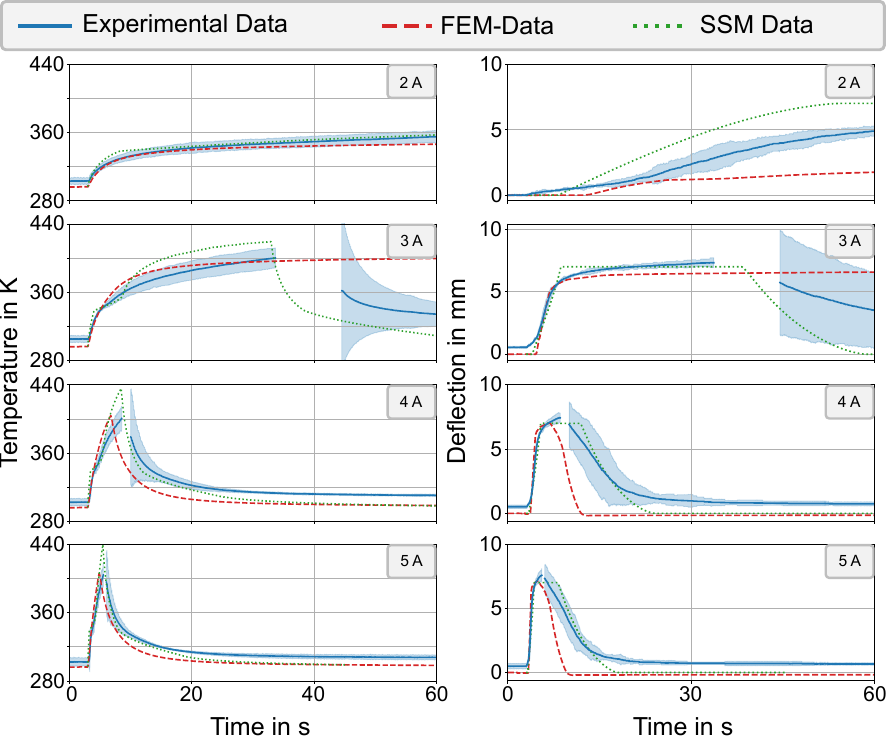}
	\caption{Temperature (left) and deflection (right) over time of an actuator Type C at room temperature under the influence of different current levels.}
	\label{fig: Auswertung_Current}
\end{figure}

In the following, FE simulation results are compared with data obtained from the SSM and with the previously published experimental results. Because the moment at which heating ended differed between specimens in the experimental investigations, the mean temperature and its confidence interval could not be computed steadily in this region. Therefore there is a gap between the end of the displayed experimental data for heating and cooling.\\
Compared to the results obtained from the SSM, the FE simulation offers the advantage that its results can also be visualized by illustrating the mesh deformation at different time steps within the calculation interval. As shown in Figure \ref{fig: Auslenkung} the temperature distribution across the entire model can be displayed at various time steps, while simultaneously providing a reference state. Therefore the temperature increase of the SMA wire due to Joule heating can also be evaluated quantitatively. As depicted in Figure \ref{fig: Auslenkung}, an inhomogeneous temperature distribution occurs along the length of the wire. The center of the wire heats up significantly faster than the end regions, until the specified shut-off temperature of 403 K is reached. This matches the experimental investigations, demonstrating the FE simulation’s ability to produce physically consistent temperature distributions.
Several FE simulations were conducted to reproduce different scenarios from the real experiments. In an initial series, the influence of different current intensities (2 A to 5 A) on the temperature increase of the SMA wire and the deflection of the SMAHC was analyzed. Figure \ref{fig: Auswertung_Current} presents both the temperature (left) and the deflection (right) as functions of time and compares the results of the FE simulation (dashed red line), the SSM (dotted green line), and the experimental results (solid blue line).
As in the experimental case, the temperature in the FE simulation was measured at the same location, namely at a node located at the center of the actuator on the outer side of the resin layer surrounding the SMA wire (see Figure \ref{fig: Auslenkung}). When comparing the temperature curves in both cases, a degressive temperature increase is observed until the shut-off is reached at 403 K, followed by a degressive cooling phase down to room temperature. During the heating phase, good agreement is achieved across all current intensities. Only at a current of 3 A the FE simulation fails to reach the shut-off criterion within the considered time interval, whereas it is attained in the experimental data. The shut-off criterion is missed by 2.8 K during the simulated period (90 s). More pronounced discrepancies arise during the cooling. The experimental data exhibits significantly slower cooling compared to the simulations and does not reach the initial ambient temperature within the observed time interval, while in the simulations it is already reached a few seconds after the current is switched off. This behavior is even more pronounced in the FE simulation compared to the SSM simulation. Furthermore, a distinct kink in the heating curve of the staggered-scheme model is not captured by the FE simulation. This kink can be attributed to the latent heat of the phase transformation associated with the change from Martensite to the Austenite phase. This is surprising since the utilized material model includes this effect and was correctly parameterized. Still, overall the observed agreement between the FE simulation and experimental data is considered reasonable.\\
When examining the deflection curves, it can be observed that, the positive deflection of the actuator lies within the confidence interval in all test series, except for 2 A. The maximum deflection values are also within plausible ranges and exhibit only minor deviations from the experimental results. Across all investigated test series, a general tendency toward slightly lower maximum deflections in the FE simulation could be observed. The deviation at 2 A may be attributed to excessive stiffness of the modeled actuator due to overestimated stiffness for the resin layer and the soft interlayer, which leads to a greater shift in the transformation temperatures of the martensitic and austenitic phases. In addition, the FE simulation appears thermally more inert at lower current intensities (2 A and 3 A) and thermally faster at higher current intensities (4 A and 5 A). However, this observation can be attributed to the statistical variation of the experimental data or the accuracy of the FE simulation. At lower currents the shut-off temperature is not reached within 60 seconds. For 2 A this is consistent with the experimental observation. At 3 A, the discrepancy between the experiment (which reaches the cut-off temperature) and the FE simulation (which does not) may be explained by the previously observed higher thermal inertia at lower current intensities. This shifts the transition between heating and cooling to a later phase of the experiment. This contributes to the observed time shift between simulated and experimental rebound curve. Furthermore, due to its slower cooling, the real actuator generally exhibits a slower return to its original position than the simulated actuator. Overall, the FE-simulated deflection curves exhibit good qualitative agreement with the experimental observations.\\
\begin{figure}[h]
	\centering
	\includegraphics[width=1\linewidth]{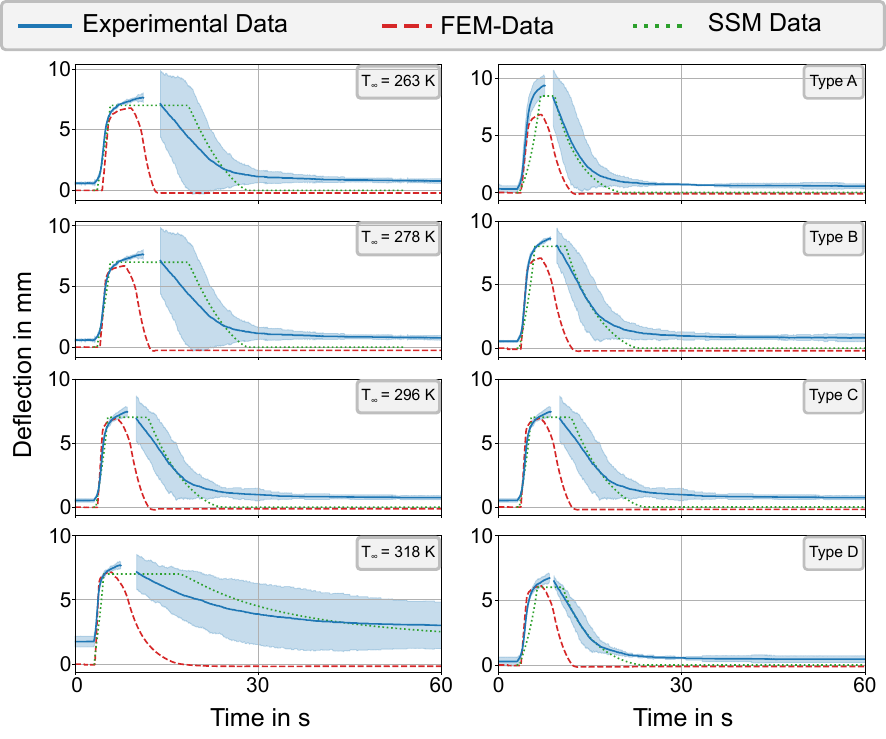}
	\caption{Left: Deflection over time of actuator Type C at a current of 4 A for different ambient temperatures.\\Right: Deflection over time of different actuator types at a current of 4 A at room temperature (296 K).}
	\label{fig: Auswertung_Tamb}
\end{figure}
In further experimental series, the influence of ambient temperature as well as geometric properties of the actuator on the deflection was investigated. The left side of Figure \ref{fig: Auswertung_Tamb} shows the deflection over time of a Type C actuator (thin elastomer) for ambient temperatures from 263 K to 318 K. In the positive deflection direction, a good agreement between FE simulation and experiment can initially be observed. As seen previously, the faster heating process in the FE simulation leads to the shut-off criterion being reached significantly earlier. Consequently, the return motion towards the neutral position begins much earlier in the FE simulation than in the experimental data. As in the previous cases, the deflection over time in the real process behaves more inert compared to the FE simulation results, which can be attributed to the slower cooling in the experiments. The largest deviation can be observed during the cooling process at an ambient temperature of 318 K. The experimental data indicates that, due to the high ambient temperature during cooling, the actuator does not return to its initial position. It is also noticeable that the initial position itself is not located at 0 mm. In contrast, the FE simulation starts at a deflection of 0 mm and, after reaching the shut-off criterion, quickly returns to this position.\\
The right side of Figure \ref{fig: Auswertung_Tamb} illustrates the deflection over time of different actuator types under identical process parameters (current: 4 A, ambient temperature: 296 K). The deflection characteristics for the different actuator types A to D, as shown in Figure \ref{fig: Auswertung_Tamb}, exhibit the same discrepancy between FE simulation and experiment during the cooling phase. However, noticeable differences between FE simulation and experiment or data from the SSM occur in the maximum deflection values. While good agreement between FE simulation and experiment can still be observed for actuator types C and D, increasing deviations appear with thinner elastomer layers in actuator types A and B. According to the experimental results, the smaller distance between SMA wire layer and substrate, that comes with a thinner elastomer layer, should lead to a higher maximum deflection. However, this effect is not reproduced in the FE simulation. A possible reason for this discrepancy is likely the linear-elastic modeling approach for the soft interlayer as well as an overestimation of its stiffness in the FE simulation and an overestimation of the mechanical influence of the resin layer coating the SMA wires.\\
\begin{figure}[h]
	\centering
	\includegraphics[width=1\linewidth]{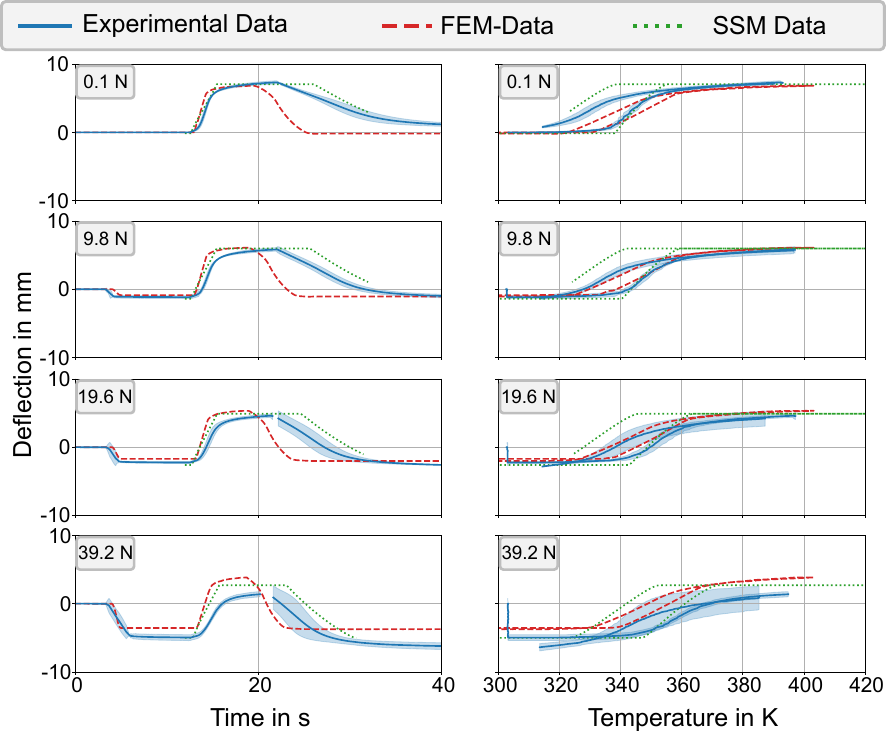}
	\caption{Left: Deflection over time of a Type C actuator at room temperature and a current of 3.5 A for various loads.\\Right: Deflection-temperature hysteresis for a Type C actuator at room temperature and a current of 3.5 A under various loads.}
	\label{fig: Auswertung_Load}
\end{figure}
Figure \ref{fig: Auswertung_Load} presents the results of a series investigating the influence of external loads applied at the free end of the actuator and their influence on its deflection. The deflection as a function of time (left) and as a function of temperature (right) are shown. The deflection–temperature hysteresis, which is characteristic for SMA materials, is clearly reproduced in the FE simulation. While the hysteresis shows good agreement in terms of width, for higher loads it is shifted to higher deflections, indicating that the SMAHC is modeled as too stiff. For higher loads, pseudoplastic deformation of the SMA wire can also be observed in the FE simulation, although it is significantly less pronounced than in the experimental results. While it can be observed that the deviations of the simulated hysteresis curves from the experimental ones increase as the load increases, the statistical variation also increases, which makes it slightly less likely that this is actually a systematic trend. For the load of 39.2 N, at the beginning of the experiment, when the actuator is deflected downwards by the applied load and while no current is applied, the deviation between simulated and measured deflection increases with increasing load, which is another manifestation of the previous observation that the FE-simulated actuator is stiffer than the physical one. Overall, the FE simulation results are in good qualitative agreement with the experimental data for loads ranging from 0.1 N to 19.6 N.

\section{Summary, Conclusions and Outlook}
\label{Conclusion}
The aim of this work was to evaluate the suitability of the recently implemented material model of \cite{kellyMicromechanicsinspiredConstitutiveModel2016} for the FE simulation of SMAHC actuators. Therefore its results were compared with those of a staggered scheme numerical model and physical experiments on the commercially available A3950 actuator by CompActive GmbH. By using the physically motivated material model \cite{kelly_micromechanics-inspired_2016}, which also takes into account the microscopic evolution of the thermoelastic phase transition this approach should for the first time be able to model the thermoelastic response of the SMA precisely in a full 3D FEM environment. The model was successfully validated using previously published data from the experimental characterization of the modeled SMAHC actuators as well as from a staggered scheme reference model \cite{kaiserExperimentallyCharacterizationTheoretical2023}. Also for the first time, the developed FE-model incorporates mechanical, thermal, and electromagnetic coupling and simulates the electro-thermomechanical response of an SMA wire as an integrated component of the overall SMAHC actuator model. The relevant output parameters for this application include the achievable deflection after thermal activation both with and without applied load. The comparison with the experimental results shows that, despite the fact that not all geometric parameters and material properties were precisely known, the FEM simulation was still capable of predicting the dynamic response of the actuator reasonably well. In total, the following conclusion could be drawn:
\begin{itemize}
    \item Joule heating of the SMA wire is modeled with qualitatively good agreement concerning heating rate and maximum temperature.
    \item Convective cooling shows significant discrepancy between experiment and FEM simulation, which is more pronounced at higher ambient temperatures.
    \item Overall, the deflection is modeled reasonably well 
    \item In general, deviations between FEM simulation and experimental data increase with higher load and higher ambient temperatures.
    \item In most cases the results from SSM show better agreement with the experimental data than the FEM simulation. 
\end{itemize}
Future work will exploit the model’s capability to compute the martensite volume fraction $\lambda$ with spatial resolution, use an improved material model for the elastomer interlayer and conduct a sensitivity analysis of material properties and geometric parameters.

\section*{Acknowledgments}
\label{Acknowledgments}

The authors would like to express their sincere gratitude to Jesper Karlsson of the DYNAMORE Nordic support team for his valuable assistance, as well as for providing a specialized solver version (ls-dyna\_smp\_d\_R16\_1213-gb48df2a0d7\_winx64\_ifort190.exe) that enabled the resolution of an issue related to the implemented SMA material model.

\newpage

\printbibliography
\newpage
\appendix 

\section{Actuator Geometry}
\begin{table}[hb]
\caption{\label{GeoParameter}Geometric parameters of actuator modules of types A, B, C, and D according to \cite{kaiserExperimentallyCharacterizationTheoretical2023}. The geometric parameters vary only where the types are explicitly mentioned.}
\footnotesize
\begin{tabular}{@{}llll}
\br
Parameter&Description & Value in mm\\
\mr
Active wire length                                                   &      l\textsubscript{active}       & 38.65      \\
SMA wire length                                                      &      l\textsubscript{Wire}       & 42.61      \\
scaled wire length                                                   &      -       & 41.6191    \\
Wire diameter                                                        &      d\textsubscript{Wire}       & 0.48       \\
Actautor width                                                       &      w\textsubscript{Actuator}       & 5.00          \\\begin{tabular}[c]{@{}c@{}}Substrate to wire\\ distance\end{tabular} &      p\textsubscript{SMA}       &  A = 1.517; B = 2.050; C = 2.677; D = 2.625\\
Substrate thickness                                                  &      t\textsubscript{Sub}       & A, B, C = 0.40; D = 0.50        \\
Rigid substrate length                                               &      l\textsubscript{Rigid}       & 22.66      \\
Pointer subtrate length                                              &      l\textsubscript{Pointer}       & 68.70      \\
Anchor clip length                                                &      l\textsubscript{Anchor}       & 1.50            \\
Anchor clip thickness                                                &      t\textsubscript{active}       & 1.22           \\
Crimpbar length                                                      &      l\textsubscript{Crimp}       & 9.00           \\
Crimpbar thickness                                                   &      t\textsubscript{Crimp}       & 1.85           \\
Tape thickness                                                       &      t\textsubscript{Tape}       & 0.25           \\
Resin thickness                                                      &      t\textsubscript{Tape}& 0.20           \\
\br
\end{tabular}
\end{table}
\newpage
\section{Material Parameters}

\begin{table}[hb]
\caption{\label{MechParameterSMA}Structural-mechanical and thermal material parameters of the SMA.}
\footnotesize
\begin{tabular}{@{}lllll}
\br
Parameter & Description & Value & Unit & Source\\
\mr
RHO   & \begin{tabular}[c]{@{}c@{}}Density of the\\ SMA\end{tabular} & $6.561\cdot 10^{-6}$ & $\frac{kg}{mm^{3}}$ & \cite{kaiserExperimentallyCharacterizationTheoretical2023}\\
EM   & \begin{tabular}[c]{@{}c@{}}E-Modulus\\ Martensite\end{tabular} & 19.17 & $GPa$ & \cite{kaiserExperimentallyCharacterizationTheoretical2023}\\
EA   & \begin{tabular}[c]{@{}c@{}}E-Modulus\\ Austenite\end{tabular} & 30.25 & $GPa$ & \cite{kaiserExperimentallyCharacterizationTheoretical2023}\\
PRM   & \begin{tabular}[c]{@{}c@{}}Poisson's ratio\\ Martensite\end{tabular} & 0.468 & - & \cite{karlssonMAT_291NewMicromechanicsinspired2019}\\
PRA   & \begin{tabular}[c]{@{}c@{}}Poisson's ratio\\ Austenite\end{tabular} & 0.45 & - & \cite{karlssonMAT_291NewMicromechanicsinspired2019}\\
CPM   & \begin{tabular}[c]{@{}c@{}}Volumetric heat capacity\\ Martensite\end{tabular} & $1.915\cdot 10^{-3}$ & $\frac{J}{mm^{3}K}$ & \cite{kaiserExperimentallyCharacterizationTheoretical2023}\\
CPA   & \begin{tabular}[c]{@{}c@{}}Volumetric heat capacity\\ Austenite\end{tabular} & $3.206\cdot 10^{-3}$ & $\frac{J}{mm^{3}K}$ & \cite{kaiserExperimentallyCharacterizationTheoretical2023}\\
LH   & \begin{tabular}[c]{@{}c@{}}Volumetric latent heat\\ of the transformation\end{tabular} & 0.123 & $\frac{J}{mm^{3}}$ & \cite{kaiserExperimentallyCharacterizationTheoretical2023}\\
TC   & Thermodynamic temperature & 0.0 & $K$ & \cite{karlssonMAT_291NewMicromechanicsinspired2019}\\
TMF   & Martensite finish temperature & 304.012 & $K$ & \cite{kaiserExperimentallyCharacterizationTheoretical2023}\\
TMS   & Martensite start temperature & 327.77 & $K$ & \cite{kaiserExperimentallyCharacterizationTheoretical2023}\\
TAS   & Austenite start temperature & 341.49 & $K$ & \cite{kaiserExperimentallyCharacterizationTheoretical2023}\\
TAF   & Austenite finish temperature & 359.27 & $K$ & \cite{kaiserExperimentallyCharacterizationTheoretical2023}\\
D0M   & \begin{tabular}[c]{@{}c@{}}Initial driving force for \\ martensite strain transformation\end{tabular} & 0.05 & $Gpa$ & \cite{karlssonMAT_291NewMicromechanicsinspired2019}\\
D0L   & \begin{tabular}[c]{@{}c@{}}Initial driving force for \\ volume fraction transformation\end{tabular} & 0.05 & $Gpa$ & \cite{karlssonMAT_291NewMicromechanicsinspired2019}\\
SIGMA   & Electrical conductivity & $1.22\cdot 10^{6}$ & $\frac{1}{k\Omega mm}$ & \cite{kaiserExperimentallyCharacterizationTheoretical2023}\\
\br
\end{tabular}
\end{table}
\newpage
\begin{table}[t!]
\caption{\label{MechParameter}Structural-mechanical material parameters of the individual components.}
\footnotesize
\begin{tabular}{@{}lllll}
\br
Components&Material &Density ($\frac{kg}{mm^{3}}$)&E-Modulus ($GPa$)&Poissons's ratio\\
\mr
Substrate   & Steel                                                             &    $7.876\cdot10^{-6}$    &    $215.67$      & $0.42$                     \\
Interlayer & Elastomer                                                         &    $1.038\cdot10^{-6}$   &    $8.53$      & 0.29                     \\
Anchorclip & \begin{tabular}[c]{@{}c@{}}Thermoplastic\\ polymer\end{tabular} &    $1.27\cdot10^{-6}$    &    $3.2$      & $0.44$                     \\
Crimpbar   & Brass                                                           &    $8.46\cdot10^{-6}$    &    $96$      & $0.34$                     \\
Resinlayer & Resin                                                    &    $1.17\cdot10^{-6}$    &    $3.454$      &$0.35$\\
\br
\end{tabular}
\end{table}
\FloatBarrier
\begin{table}[H]
\caption{\label{ThermalParameter}Thermal material parameters of the individual components}
\footnotesize
\begin{tabular}{@{}llll}
\br
Components&Material &Specific heat capacity ($\frac{kN\space mm}{kg\space K}$)&Thermal conductivity ($\frac{W}{m\space K}$)\\
\mr
Substrate   & Steel                                                             &          $484$            &            $15$           \\
Interlayer & Elastomer                                                         &          $1619$            &            $0.18$          \\
Anchorclip & \begin{tabular}[c]{@{}c@{}}Thermoplastic\\ polymer\end{tabular} &          $1200$            &            $0.24$          \\
Crimpbar   & Brass                                                           &          $377$            &            $113$          \\
Resinlayer & Resin                                                    &          $1289$            &            $0.2$          \\
\br
\end{tabular}
\end{table}
\end{doublespace}
\end{document}